\documentclass[11pt]{article}
\usepackage{graphicx,color}  
\usepackage{dcolumn}   
\usepackage{bm}        
\usepackage{amssymb} 
\usepackage{amsmath}
\usepackage{braket}
\usepackage{cancel}
\usepackage{subfigure}
\usepackage{fullpage}

\usepackage{tcolorbox}
\usepackage{tikz}
\usepackage{pgf}
\usetikzlibrary{arrows}
\usetikzlibrary{snakes}
\usetikzlibrary{decorations,decorations.pathmorphing,decorations.markings}
\usepackage{fullpage}

\tikzset{graviton/.style={decorate, decoration={snake}, double}}

\usepackage{adjustbox}
\usepackage{standalone}
\usepackage{tikz}
\usepackage{tikz-feynman}
\usepackage{pgf}
\usepackage{cancel}
\usepackage{braket}
\usepackage[T1]{fontenc}
\usepackage[utf8]{inputenc}
\usepackage{xcolor}
\usepackage{tikz-feynman}
\usepackage[italian,english]{babel}
\usepackage{url}
\usepackage{siunitx}
\usepackage{amsmath}
\usepackage{amssymb}
\usepackage{graphicx}
\usepackage{graphicx,caption}
\usepackage{siunitx}
\usepackage{bm}
\usepackage{amsfonts}
\usepackage[left=2cm,right=2cm,top=2.5cm,bottom=2.5cm]{geometry}
\usepackage{amsmath}
\usepackage{graphicx}
\usepackage{subfigure}
\usepackage{slashed}
\usepackage{verbatim}

\newcommand{\beq}{\begin{equation}}
	\newcommand{\eeq}{\end{equation}}
\newcommand{\beqa}{\begin{eqnarray}}
	\newcommand{\eeqa}{\end{eqnarray}}
\newcommand{\be}{\begin{equation}}
	\newcommand{\ee}{\end{equation}}

\newcommand{\nc}{\newcommand}
\nc{\rnc}{\renewcommand}
\rnc{\d}{\mathrm{d}}
\nc{\D}{\partial}
\nc{\K}{\kappa}
\nc{\bK}{\bar{\K}}
\nc{\bN}{\bar{N}}
\nc{\bq}{\bar{q}}
\nc{\vbq}{\vec{\bar{q}}}
\nc{\g}{\gamma}
\nc{\lrarrow}{\leftrightarrow}
\nc{\rg}{\sqrt{g}}
\nc{\nn}{\nonumber}
\rnc{\(}{\left(}
\rnc{\)}{\right)}
\nc{\q}{\vec{q}}
\nc{\x}{\vec{x}}
\rnc{\a}{\hat{a}}
\nc{\ep}{\epsilon}
\nc{\tto}{\rightarrow}
\rnc{\inf}{\infty}
\rnc{\Re}{\mathrm{Re}}
\rnc{\Im}{\mathrm{Im}}
\nc{\z}{\zeta}
\nc{\mA}{\mathcal{A}}
\nc{\mB}{\mathcal{B}}
\nc{\mC}{\mathcal{C}}
\nc{\mD}{\mathcal{D}}
\nc{\mN}{\mathcal{N}}
\rnc{\H}{\mathcal{H}}
\rnc{\L}{\mathcal{L}}
\nc{\<}{\langle}
\rnc{\>}{\rangle}
\nc{\fnl}{f_{NL}}
\nc{\gnl}{g_{NL}}
\nc{\fnleq}{f_{NL}^{equil.}}
\nc{\fnlloc}{f_{NL}^{local}}
\nc{\vphi}{\varphi}
\nc{\Lie}{\pounds}
\nc{\half}{\frac{1}{2}}
\nc{\bOmega}{\bar{\Omega}}
\nc{\bLambda}{\bar{\Lambda}}
\nc{\dN}{\delta N}
\nc{\gYM}{g_{\mathrm{YM}}}
\nc{\geff}{g_{\mathrm{eff}}}
\nc{\tr}{\mathrm{tr}}
\nc{\oa}{\stackrel{\leftrightarrow}}
\nc{\IR}{{\rm IR}}
\nc{\UV}{{\rm UV}}
\nc{\la}{\lambda}
\nc{\veps}{\varepsilon}
\begin{document}
	\begin{center}
		\vspace{1.5cm}
		{\Large \bf  CFT Constraints on Parity-odd Interactions\\ with Axions and Dilatons\\}

		\vspace{1cm}
		{\bf Claudio Corian\`o$^{1,2,3}$ and Stefano Lionetti$^{1,2}$ \\}

		\vspace{1cm}
		{\it  $^{1}$Dipartimento di Matematica e Fisica, Universit\`{a} del Salento 
			and\\  INFN Sezione di Lecce, Via Arnesano 73100 Lecce, Italy\\}
		\vspace{0.3cm}
		{\it  $^{2}$
			National Center for HPC, Big Data and Quantum Computing,\\ Via Magnanelli 2, 40033 Casalecchio di Reno, Italy\\}
		\vspace{0.3cm}
		{\it  $^{3}$  Institute of Nanotechnology, \\ National Research Council (CNR-NANOTEC), Lecce 73100\\}
		\vspace{0.5cm}

		\begin{abstract}
We illustrate how the conformal Ward identities (CWIs) in momentum space completely determine the structure of a parity-odd 3-point correlator involving currents, energy-momentum tensors and at least one scalar operator in $d=4$. Conformal invariance fixes almost all such possible correlators to vanish. The only exceptions are given by the $\langle JJO\rangle_{odd}$  and the $\langle TTO \rangle_{odd}$ which in momentum space are protected by chiral and conformal anomalies. Specifically, one can obtain a non-vanishing solution by considering scalar operators such as $O=\nabla \cdot J_A$, $O=g_{\mu\nu}T^{\mu\nu}$ or their shadow transforms. We comment on the implications of these results that constrain the coupling of axions and dilatons in a conformal phase of the early universe. 		
\end{abstract}
	\end{center}
\newpage
\section{Introduction}
For over half a century, Conformal Field Theories (CFTs) have been instrumental in numerous areas of theoretical physics, ranging from the theory of critical phenomena to string theory and the AdS/CFT correspondence. Conformal invariance imposes stringent constraints on the structure of correlation functions. The general forms of two- and three-point correlation functions are determined by conformal symmetry up to a finite set of constants and these forms can be derived via coordinate space techniques. However, such techniques are confined to configurations where the points in the correlators are distinct. They provide limited insights into the origins of quantum anomalies manifesting at short distances, where points coalesce. Consequently, one key motivation for analyzing CFT correlation functions directly in momentum space is to elucidate the impact of anomalies more transparently. Moreover, given that perturbative field theory is naturally articulated in momentum space, examining CFTs in this framework is advantageous. \\
Determining correlation functions in momentum space is, however, significantly more intricate than in position space. The use of conformal Ward identities to determine the structure of three-point functions in momentum space was independently introduced in \cite{Coriano:2013jba} and \cite{Bzowski:2013sza}, with the latter extending the methodology to tensor cases. Renormalized parity even correlators, type-A and type-B Weyl anomalies have also been extensively discussed  \cite{Bzowski:2015pba,Bzowski:2018fql,Bzowski:2017poo,Coriano:2020ees}. Furthermore, previous perturbative analyses in QED and QCD have examined important correlators such as the parity-even $\langle TJJ \rangle$, involving the energy-momentum tensor $T$ and currents $J$, in both conformal and non-conformal settings \cite{Giannotti:2008cv,Armillis:2009pq,Armillis:2010qk}. \\
Recently, parity-odd correlators and non-conserved currents have also been the focus of attention \cite{Jain:2021gwa,Jain:2021wyn,Buchbinder:2023coi,Marotta:2022jrp}. In particular, in 
\cite{Coriano:2023hts, Coriano:2023cvf,Coriano:2023gxa} parity-odd correlators have been explored in relation to chiral and conformal anomalies.
The latter papers demonstrate how such anomalies are responsible for generating non-vanishing parity-odd correlators. In the absence of chiral anomalies, correlators like $\langle J_V J_V J_A \rangle$, $\langle J_A J_A J_A \rangle$, $\langle TTJ_A \rangle$ — where $J_V$ and $J_A$ are vector and axial-vector currents respectively —  would entirely vanish in a CFT. Likewise, without parity-odd conformal anomalies, the $\langle TJJ\rangle_{odd} $ correlator, where $J$ is a generic current of any parity, is zero.\\
The recent applications of these results to topological insulators and Weyl semimetals has rejuvenated interest in this domain and in several dimensions \cite{Landsteiner:2022nbl,Arouca:2022psl,Coriano:2024ive}.
Parity-odd structures arise naturally in various contexts, including CFTs with broken parity symmetry such as Chern-Simons matter theories. Another significant application is in cosmology, where the bispectrum of the Cosmic Microwave Background (CMB), which is a measure of non-gaussianity, is given by the three-point function in momentum space \cite{Lue:1998mq,Maldacena:2011nz,Coriano:2012hd, Soda:2011am,Shiraishi:2014roa,Shiraishi:2014ila}.\\
In three dimensions, the computation of the correlators simplifies considerably due to the absence of anomalies. However, it remains significantly important, particularly in the context of the AdS/CFT correspondence \cite{Raju:2012zs}. Using a holographic approach, these correlators describe the curvature gravitational perturbations in a pre-inflationary phase of the early universe characterized by strong gravity \cite{McFadden:2011kk, Bzowski:2011ab, Maldacena:2011nz,Afshordi:2016}. They are crucial in studying the non-Gaussian contributions to such perturbations at the level of the bispectrum (via the $\langle TTT\rangle$ correlator) and the trispectrum (via the $\langle  TTTT\rangle$ correlator), while the power spectrum is determined by the $\langle TT\rangle$ correlator \cite{McFadden:2009fg,Coriano:2021}.\\
In this paper we further examine the conformal constraints for parity-odd correlators in momentum space in $d=4$. We focus in particular on all the possible 3-point correlators built with currents, energy-momentum tensors and at least one scalar operator.\\
Parity-odd correlators can be realized, for example, by considering a loop interaction with chiral fermions or by directly inserting a $\gamma_5$ into an operator, as in the case of an axial fermionic current $J_A^\mu = \bar{\psi}\gamma^\mu \gamma_5 \psi$. In this paper, we study parity-odd correlators independently of their possible realizations. For example, the $\langle TJO \rangle_{odd}$ can be realized using a pseudoscalar operator, an axial current, a pseudo-tensor, or all of them simultaneously. They play a role in non-perturbative analysis of interactions between pseudoscalar fields - for example an axion-like field - and gravity in a conformal symmetric phase. 
We will comment on the possible implications of this analysis from the phenomenological point of view, before our conclusions. \\
The nature of the pseudoscalar operator introduced in our investigation is general, and can be of any scaling dimension $\Delta$. The most interesting cases, from our perspective, are for $\Delta = 4$, where $O \sim \partial J_5 $ and $ O \sim T^\mu_\mu $, which carry direct physical implications.\\
In the case of the $ \langle JJO\rangle_{odd}$, for example, our results remain identical even if $O $ and $ J $ are both of odd parity, as the behavior of the correlator under parity is a global property of the correlator and not necessarily of a single operator in the triple.
As we will show, the solution of the conformal and parity constraints naturally requires that their tensor structure is essentially that induced by a chiral anomaly. A modification of this behavior is encountered in the shadow solutions of the CWIs, where a primary operator $ O $ of scaling dimension $ \Delta $ is replaced by its shadow-transformed expression of dimension $ d - \Delta $. The corresponding operators, in this case, can be defined by a non-local shadow transform. \\
Our work is organized as follows. In Section \ref{secCWIS}, we describe the CWIs in coordinate and in momentum space. In Section \ref{sec2}, we briefly discuss parity-odd correlators with two and three scalar operators. In Section \ref{sec3}, we then derive the solution of the CWIs for the $\langle TJO\rangle_{odd}$. This correlator interpolates with the (parity-odd) amplitude for an axion/dilaton particle transitioning into a spin-1 state in the presence of an off-shell external gravitational field. As we will demonstrate, this amplitude vanishes identically in a conformal phase of the early universe. Then, in Section \ref{sec4}, we turn to reinvestigate the $\langle JJO\rangle_{odd} $ and the $\langle TTO\rangle_{odd} $ correlator. For the latter, we show that the CWIs admit additional solutions that had not been identified in a previous analysis \cite{Coriano:2023cvf}. Such solutions are connected by shadow transforms of the primary scalars. 
Lastly, before presenting our conclusions, in Section \ref{sec5}, we discuss the relevance of this study at the cosmological level, specifically regarding the analysis of chiral and dilatation current anomalies in the early universe.

\section{Conformal Ward identities from coordinate to momentum space}
\label{secCWIS}
The Poincaré group is a fundamental symmetry group in physics, encompassing translations and Lorentz transformations. These transformations leave the metric invariant, making the Poincaré group essential in the study of special relativity and quantum field theory. However, certain physical theories, such as those involving massless particles (like photons) or in the study of critical phenomena, exhibit a broader set of symmetries. This broader symmetry group is known as the conformal group, which extends the Poincaré group to include additional transformations that preserve the angle between spacetime intervals, rather than their exact length.
In addition to the Poincaré transformations, the conformal group includes dilations and special conformal transformations. This extension is crucial for understanding systems with scale invariance, such as massless field theories, critical phenomena, and the geometry of spacetime in the context of the AdS/CFT correspondence.  \\
Correlators that are invariant under the full conformal group need to satisfy peculiar relations that highly constrain their structure. Let us consider a scalar $n$-point correlator function $\left\langle{O}_1\left(x_1\right) \ldots {O}_n\left(x_n\right)\right\rangle$  where $\Delta_1,\dots,\Delta_n$ represent the dimensions of the operators.
The Ward identity for an infinitesimal transformation $g$ can be expressed as
\begin{equation}
	0=\sum_{j=1}^n\braket{{O}_1(x_1)\dots\delta_g{O}_j(x_j) \dots{O}_n(x_n)},\label{globalW}
\end{equation} 
We now recall the conformal transformations rules
\begin{equation}
	\begin{split}
		[P_\mu,{O}_j(x)]&=\partial_\mu{O}_j(x),\\
		[D,{O}_j(x)]&=(\Delta_{j}+x^\mu\partial_\mu){O}_j(x),\\
		[L_{\mu\nu},{O}_j(x)]&=\left(\mathcal{S}_{\mu\nu}+x_\nu\partial_\mu-x_\mu\partial_\nu\right){O}_j(x),\\
		[K_\mu,{O}_j(x)]&=\left[2x_\mu (x\cdot\partial)-x^2\partial_\mu+2\Delta_{j} x_\mu-2x^\nu\mathcal{S}_{\mu\nu}\right]{O}_j(x).
	\end{split}\label{transform}
\end{equation}
Here, $\mathcal{S}_{\mu\nu}$ is a finite-dimensional representation matrix of rotations, which determines the spin of the field and vanishes in the case of a scalar operator.
The dilatation Ward identity can then be expressed as
\begin{equation}
	0=\left[\sum_{j=1}^n \Delta_j+\sum_{j=1}^n x_j^\alpha \frac{\partial}{\partial x_j^\alpha}\right]\left\langle{O}_1\left(x_1\right) \ldots {O}_n\left(x_n\right)\right\rangle
\end{equation}
while the special conformal Ward identity is given by
\begin{equation}\label{eq:specconfwicoord}
	0=\sum_{j=1}^n\left[2 \Delta_j x_j^k+2 x_j^\kappa x_j^\alpha \frac{\partial}{\partial x_{j\, \alpha}}-x_j^2 \frac{\partial}{\partial x_{j k}}\right]\left\langle{O}_1\left(x_1\right) \ldots {O}_n\left(x_n\right)\right\rangle
\end{equation}
where $k$ is a free Lorentz index. In the case of tensor operators one needs to include an additional term
to the previous equation, i.e. the contribution $\mathcal{S}_{\mu\nu}$ in eq$.$ \eqref{transform}.
For example, the special CWI of the $\langle TJO\rangle$ correlator takes the following form
\begin{equation}
	\begin{aligned}
		0=&\sum_{j=1}^3\left[2 \Delta_j x_j^k+2 x_j^\kappa x_j^\alpha \frac{\partial}{\partial x_{j\, \alpha}}-x_j^2 \frac{\partial}{\partial x_{j k}}\right]\left\langle  T^{\mu_1 \nu_1}\left(x_1\right) J^{\mu_2}\left(x_2\right)O\left(x_3\right)\right\rangle\\ 
		&+4\bigg( (x_1)_\alpha \delta^{\kappa\left(\mu_1\right.}-  \delta_{\alpha}^\kappa\, x_1^{\left(\mu_1\right.} \bigg) \langle T^{\left.\nu_1\right) \alpha}\left(x_1\right) J^{ \mu_2}\left(x_2\right) O	\left(x_3\right)\rangle  \\
		&  +2\bigg((x_2)_{\alpha} \delta^{\kappa\mu_2} -\delta_{\alpha}^\kappa \, x_2^{\mu_2}   \bigg)\left\langle  T^{\mu_1 \nu_1}\left(x_1\right) J^{\alpha}\left(x_2\right)O\left(x_3\right)\right\rangle .
	\end{aligned}
\end{equation}
The CWIs above may be Fourier transformed in a similar manner to that discussed in \cite{Coriano:2018bbe}. Due to the translation invariance the position space correlators depend only on the differences $x_j- x_n$. Therefore, we can set $x_n = 0$ and take
\begin{equation}
	{p}_n=-\sum_{j=1}^{n-1} {p}_j
\end{equation}
The dilatation equation in momentum space takes the form
\begin{equation}
	\left[\sum_{j=1}^n \Delta_j-(n-1) d-\sum_{j=1}^{n-1} p_j^\alpha \frac{\partial}{\partial p_j^\alpha}\right] 
	\left\langle{O}_1\left(p_1\right) \ldots {O}_n\left(p_n\right)\right\rangle
	=0
\end{equation}
while the special conformal Ward identity is given by
\begin{equation}
	\begin{aligned}
		0=\sum_{j=1}^{n-1}\left(2\left(\Delta_j-d\right) \frac{\partial}{\partial p_j^\kappa}-2 p_j^\alpha \frac{\partial}{\partial p_j^\alpha} \frac{\partial}{\partial p_j^\kappa}+\left(p_j\right)_\kappa \frac{\partial}{\partial p_j^\alpha} \frac{\partial}{\partial p_{j \alpha}}\right)
		\left\langle{O}_1\left(p_1\right) \ldots {O}_n\left(p_n\right)\right\rangle
	\end{aligned}
\end{equation}
Once again, in the tensorial case we need to include an additional term on the right-hand side of the special conformal equation. For example, in the case of the $\langle TJO\rangle$, the Ward identity takes the following form
\begin{equation}
	\begin{aligned}
		0=& \sum_{j=1}^2\left(2\left(\Delta_j-d\right) \frac{\partial}{\partial p_{j \kappa}}-2 p_j^\alpha \frac{\partial}{\partial p_j^\alpha} \frac{\partial}{\partial p_{j \kappa}}+\left(p_j\right)^\kappa \frac{\partial}{\partial p_j^\alpha} \frac{\partial}{\partial p_{j \alpha}}\right)\left\langle T^{ \mu_1\nu_1}\left(p_1\right) 
		J^{ \mu_2}\left(p_2\right) O	\left(p_3\right)\right\rangle \\
		& +4\left(\delta^{\kappa\left(\mu_1\right.} \frac{\partial}{\partial p_1^{\alpha_1}}-\delta_{\alpha_1}^\kappa \delta_\lambda^{\left(\mu_1\right.} \frac{\partial}{\partial p_{1 \lambda}}\right) \langle T^{\left.\nu_1\right) \alpha_1}\left(p_1\right) J^{ \mu_2}\left(p_2\right) O	\left(p_3\right)\rangle  \\
		& +2\left(\delta^{\kappa\mu_2} \frac{\partial}{\partial p_2^{\alpha_2}}-\delta_{\alpha_2}^\kappa \delta_\lambda^{\mu_2} \frac{\partial}{\partial p_{2 \lambda}}\right)\left\langle  T^{\mu_1 \nu_1}\left(p_1\right) J^{\alpha_2}\left(p_2\right)O\left(p_3\right)\right\rangle .
	\end{aligned}
\end{equation}
In the following sections, we will explore how the CWIs outlined above constrain parity-odd 3-point correlators involving at least one scalar operator. We will begin by decomposing the correlator into a longitudinal-trace and a transverse-traceless sector. Subsequently, we will solve the conformal equations to determine the most general structure for a correlator in a CFT, following the methodology introduced in \cite{Bzowski:2013sza}.
As we will demonstrate, the solution to the CWIs in momentum space can be expressed using a specific class of functions known as 3K integrals, which involve products of Bessel functions.

\section{Three-point functions with multiple scalars}
\label{sec2}
We start by considering correlators with at least two scalar operators $O(p_i)$. 
Before examining the conformal constraints, we write the most general form of the correlators in terms of form factors and tensorial structures. In this respect, the vanishing of such correlators in the parity-odd sector is rather straightforward, due to symmetry. Indeed, since we are dealing with four-dimensional parity-odd correlators, the tensorial structures need to include an $\varepsilon^{\alpha_1 \alpha_2 \alpha_3 \alpha_4}$ tensor. It is easy to figure out that this condition cannot be satisfied in this case, thereby giving the vanishing relations  
\begin{equation}\label{eq:23scalop}
	\begin{aligned}
		&\left\langle O\left(p_1\right) 
		O\left(p_2\right) O	\left(p_3\right)\right\rangle_{odd}=0,\\&
		\left\langle J^{\mu}\left(p_1\right) 
		O\left(p_2\right) O	\left(p_3\right)\right\rangle_{odd}=0,\\&
		\left\langle T^{\mu \nu}\left(p_1\right) 
		O\left(p_2\right) O	\left(p_3\right)\right\rangle_{odd}=0.
	\end{aligned}
\end{equation}
We emphasize that these equations are valid without imposing full conformal invariance and even when considering non-conserved currents and energy-momentum tensors. All possible two-point functions involving scalar operators, currents, and energy-momentum tensors are zero in 4d for the same reason.\\
Conversely, if we consider a three-point correlator with only one scalar operator, it is possible to construct at least one parity-odd tensorial structure and the correlator does not necessarily vanish. In the following sections, we will examine how conformal invariance constrains such correlators by a direct application of the full methodology of CFT in momentum space.

\section{The conformal $\langle TJO\rangle_{odd}$}
\label{sec3}
In this section we concentrate on the $\langle TJO\rangle_{odd}$ correlator. From now on, we will drop the index "odd" for simplicity.\\
It is quite immediate to realize that the correlator cannot exhibit any anomaly content and its longitudinal/trace part vanishes
 \begin{equation}\label{eq:witjo}
 	\begin{aligned}
 		&0=\delta_{\mu_1\nu_1}\left\langle T^{ \mu_1\nu_1}\left(p_1\right) 
 		J^{ \mu_2}\left(p_2\right) O	\left(p_3\right)\right\rangle,	\\&0=
 		p_{1\,\mu_1}\left\langle T^{ \mu_1\nu_1}\left(p_1\right) 
 		J^{ \mu_2}\left(p_2\right) O	\left(p_3\right)\right\rangle ,
 		\\&0=
 		p_{2\,\mu_2}\left\langle T^{ \mu_1\nu_1}\left(p_1\right) 
 		J^{ \mu_2}\left(p_2\right) O	\left(p_3\right)\right\rangle .
 	\end{aligned}
 \end{equation}
Indeed, the absence of a mixed anomaly in this correlator - specifically, the lack of contractions involving the Riemann or Weyl tensors together with the field strength of the Abelian current $J$ at scaling dimension four - will be crucial for determining this correlation function.
As we are going to find out, the result of this procedure in the parity-odd sector of a generic 3-point function, is directly linked with the presence or absence of chiral anomalies. \\
In the momentum space approach, as usual, we then introduce the transverse ($\pi$) and transverse-traceless ($\Pi$) projectors
\begin{equation}
		\begin{aligned}
		\pi_\alpha^\mu=\delta_\alpha^\mu-\frac{p^\mu p_\alpha}{p^2}, \qquad \qquad \Pi_{\alpha \beta}^{\mu \nu}=\pi_\alpha^{(\mu} \pi_\beta^{\nu)}-\frac{1}{d-1} \pi^{\mu \nu} \pi_{\alpha \beta}
	\end{aligned}
\end{equation}
in order to separate its several (longitudinal, trace and transverse-traceless) sectors, expanded with respect to the external momenta. Because of eq$.$ \eqref{eq:witjo}, the correlator is composed only of a transverse-traceless part which can be written as
\begin{equation}\label{eq:TJODecomp}
	\left\langle T^{ \mu_1\nu_1}\left(p_1\right) 
	J^{ \mu_2}\left(p_2\right) O	\left(p_3\right)\right\rangle =\Pi^{\mu_1\nu_1}_{\alpha_1\beta_1}\left(p_1\right)
	\pi^{\mu_2}_{\alpha_2} \left(p_2\right) \,X^{\alpha_1\beta_1\alpha_2}.
\end{equation}
Here, $X^{\alpha_1\beta_1\alpha_2}$ is a parity-odd tensor that can be expressed in its most general form as 
\begin{equation}\label{eq:XTJO}
	X^{\alpha_1\beta_1\alpha_2}=  A(p_1,p_2,p_3)\varepsilon^{p_1 p_2 \alpha_1\alpha_2 }p^{\beta_1}_2
\end{equation}
where $\varepsilon^{ p_1 p_2\alpha_1\alpha_2}\equiv \varepsilon^{ \rho \sigma\alpha_1\alpha_2}p_{1\rho} p_{2\sigma}$ and
$A(p_1,p_2,p_3)$ is an arbitrary form factor. Therefore, the entire analysis of the correlator can be reduced to determining $A(p_1, p_2, p_3)$. In the following, we will examine how the conformal constraints fix such form factor. These constraints will take the form of differential equations for $A(p_1, p_2, p_3)$.
Specifically, we are going to solve the constraints from both the dilatation and the special conformal transformations on the parity-odd structure \eqref{eq:TJODecomp} admitted by Lorentz covariance. The special conformal equations will be separated into second order (primary) and first order (secondary) sets. The primary equations will be solved in terms of 3K integrals multiplied by an arbitrary constant. Such constant will then be set to zero by the secondary equations.

\subsection{Dilatations and special conformal equations}
We denote by $\Delta_i$ the conformal dimension of the operators in our correlator. Specifically, since we are working in four-dimensional space-time, the conformal dimensions of the energy-momentum tensor and a conserved/axial current are, respectively,
\begin{equation} \label{eq:confdimension}
	\Delta_1=4, \qquad \Delta_2=3.
\end{equation}
The invariance of the correlator under dilatation in momentum space is reflected in the equation
\begin{equation}
	\left(\sum_{i=1}^3 \Delta_i-2 d-\sum_{i=1}^2 p_i^\mu \frac{\partial}{\partial p_i^\mu}\right)\left\langle T^{\mu_1 \nu_1}\left(p_1\right) 
	J^{\mu_2}\left(p_2\right) O	\left(p_3\right)\right\rangle=0.
\end{equation} 
By using the chain rule
\begin{align}
	\frac{\partial}{\partial p_i^\mu}=\sum_{j=1}^3\frac{\partial p_j}{\partial p_i^\mu}\frac{\partial}{\partial p_j}
\end{align}
we can then express the derivatives with respect to 4-vectors in term of the invariants $p_i=|\sqrt{p_i^2}|$.\\
Furthermore, using eq$.$ \eqref{eq:TJODecomp}, \eqref{eq:XTJO} and \eqref{eq:confdimension}, we can rewrite the dilatations equations as a constraint on the form factor 
\begin{equation}\label{eq:dilatAconst}
	\sum_{i=1}^{3} p_i \frac{\partial A}{\partial p_i }(p_1, p_2, p_3)+\left(4-\Delta_3\right) A(p_1, p_2, p_3)=0.
\end{equation}
On the other hand, the invariance of the correlator under special conformal transformations is encoded in the following equation
\begin{equation}
	\begin{aligned}
		0=\mathcal{K}^\kappa & \left\langle T^{ \mu_1\nu_1}\left(p_1\right) 
		J^{ \mu_2}\left(p_2\right) O	\left(p_3\right)\right\rangle \\
		\equiv & \sum_{j=1}^2\left(2\left(\Delta_j-d\right) \frac{\partial}{\partial p_{j \kappa}}-2 p_j^\alpha \frac{\partial}{\partial p_j^\alpha} \frac{\partial}{\partial p_{j \kappa}}+\left(p_j\right)^\kappa \frac{\partial}{\partial p_j^\alpha} \frac{\partial}{\partial p_{j \alpha}}\right)\left\langle T^{ \mu_1\nu_1}\left(p_1\right) 
		J^{ \mu_2}\left(p_2\right) O	\left(p_3\right)\right\rangle \\
		& +4\left(\delta^{\kappa\left(\mu_1\right.} \frac{\partial}{\partial p_1^{\alpha_1}}-\delta_{\alpha_1}^\kappa \delta_\lambda^{\left(\mu_1\right.} \frac{\partial}{\partial p_{1 \lambda}}\right) \langle T^{\left.\nu_1\right) \alpha_1}\left(p_1\right) J^{ \mu_2}\left(p_2\right) O	\left(p_3\right)\rangle  \\
		& +2\left(\delta^{\kappa\mu_2} \frac{\partial}{\partial p_2^{\alpha_2}}-\delta_{\alpha_2}^\kappa \delta_\lambda^{\mu_2} \frac{\partial}{\partial p_{2 \lambda}}\right)\left\langle  T^{\mu_1 \nu_1}\left(p_1\right) J^{\alpha_2}\left(p_2\right)O\left(p_3\right)\right\rangle .
	\end{aligned}
\end{equation}
We then perform a transverse projection on all the indices in order to identify a set of partial differential equations
\begin{equation}
	0=\Pi_{\mu_1\nu_1}^{\rho_1\sigma_1}\left(p_1\right)
	\pi_{\mu_2}^{\rho_2} \left(p_2\right) 
	\mathcal{K}^k\left\langle T^{\mu_1 \nu_1}\left(p_1\right) 
	J^{\mu_2}\left(p_2\right) O	\left(p_3\right)\right\rangle
\end{equation}
and decompose the action of the special conformal operator on the correlator in the following way
\begin{equation}\label{eq:decompnonmink}
	\begin{aligned}
		0=\, \Pi_{\mu_1\nu_1}^{\rho_1\sigma_1}\left(p_1\right)
		\pi_{\mu_2}^{\rho_2} \left(p_2\right) &\,
		\mathcal{K}^k\left\langle T^{\mu_1 \nu_1}\left(p_1\right) 
		J^{\mu_2}\left(p_2\right) O	\left(p_3\right)\right\rangle\\ =\, 
		\Pi_{\mu_1\nu_1}^{\rho_1\sigma_1}\left(p_1\right)
		\pi_{\mu_2}^{\rho_2} \left(p_2\right)& \Bigl[		
		C_1\varepsilon^{p_1 p_2\mu_1\mu_2 }p^{\nu_1}_2p_1^\kappa
		+C_2\varepsilon^{p_1 p_2\mu_1\mu_2 }p^{\nu_1}_2p_2^\kappa
		+C_3 \epsilon^{p_1 \kappa \mu_1 \mu_2 }p^{\nu_1}_2
		+{C}_4 \epsilon^{p_2 \kappa \mu_1 \mu_2 }p^{\nu_1}_2\\ &\,
		+C_5 \epsilon^{p_1 p_2 \kappa\mu_2 }p_2^{\mu_1}p^{\nu_1}_2
		+{C}_6 \epsilon^{p_1 p_2 \kappa\mu_1 }p_3^{\mu_2}p^{\nu_1}_2+
		C_7\varepsilon^{p_1 p_2\kappa \mu_1 } \delta^{\mu_2\nu_1}
		+C_8 \varepsilon^{p_1 p_2 \mu_1 \mu_2} \delta^{\kappa \nu_1}
		\Bigr]
	\end{aligned}
\end{equation}
where $C_i$ are scalar functions that depend on the form factor $A$ and its derivatives with respect to the momenta.
The tensor structures listed in the equation above are not all independent and can be simplified in order to find a minimal decomposition, using the following Schouten identities
\begin{equation}
	\begin{aligned}
		0&=\epsilon^{[p_1 p_2 \mu_1 \mu_2} p_1^{\kappa]},\\
		0&=\epsilon^{[p_1 p_2 \mu_1 \mu_2} p_2^{\kappa]},\\
		0&=\epsilon^{[p_1 p_2 \mu_1 \mu_2} \delta^{\kappa]\nu_1},
	\end{aligned}
\end{equation}
where the square brackets indicate the antisymmetrization with respect to the enclosed indices.
Using such identities, we can eliminate the tensorial structures corresponding to $C_3$, ${C}_5$ and $C_6$
\begin{equation}
	\begin{aligned} 
		\epsilon^{p_1 p_2 {\kappa}{\mu_1}} p_3^{\mu_2}&=\frac{1}{2} \epsilon^{p_1{\kappa}{\mu_1}{\mu_2}} (p_1^2+p_2^2-p_3^2)+\epsilon^{p_1p_2{\mu_1}{\mu_2}} p_1^{\kappa}+\epsilon^{p_2{\kappa}{\mu_1}{\mu_2}} p_1^2,\\
		\epsilon^{p_1p_2{\kappa}{\mu_2}} p_2^{\mu_1}&=
		-\frac{1}{2} \epsilon^{p_2{\kappa}{\mu_1}{\mu_2}} (p_1^2+p_2^2-p_3^2)
		+\epsilon^{p_1p_2{\mu_1}{\mu_2}} p_2^{\kappa}
		-\epsilon^{p_1{\kappa}{\mu_1}{\mu_2}} p_2^2,\\
		\varepsilon^{\mu_1\mu_2\kappa p_1}p_2^{ \nu_1}&=
		-\varepsilon^{\kappa p_1 p_2 \mu_1}\delta^{\mu_2\nu_1}
		- \varepsilon^{p_1p_2\mu_1\mu_2}\delta^{\kappa \nu_1}.
	\end{aligned}
\end{equation}
Therefore, we can rewrite eq$.$ \eqref{eq:decompnonmink} in the minimal form
\begin{equation}
	\begin{aligned}
		0=\Pi_{\mu_1\nu_1}^{\rho_1\sigma_1}&\left(p_1\right)
		\pi_{\mu_2}^{\rho_2} \left(p_2\right) 
		\mathcal{K}^k\left\langle T^{\mu_1 \nu_1}\left(p_1\right) 
		J^{\mu_2}\left(p_2\right) O	\left(p_3\right)\right\rangle =
		\Pi_{\mu_1\nu_1}^{\rho_1\sigma_1}\left(p_1\right)
		\pi_{\mu_2}^{\rho_2} \left(p_2\right) \Bigl[
		C_1\varepsilon^{\mu_1\mu_2 p_1 p_2}p^{\nu_1}_2p_1^\kappa\\&
		+C_2\varepsilon^{\mu_1\mu_2 p_1 p_2}p^{\nu_1}_2p_2^\kappa
		+{C}_3 \epsilon^{p_2 \kappa \mu_1 \mu_2 }p^{\nu_1}_2
		+
		C_4\varepsilon^{p_1 p_2\kappa \mu_1 } \delta^{\mu_2\nu_1}
		+C_5 \varepsilon^{p_1 p_2 \mu_1 \mu_2} \delta^{\kappa \nu_1}
		\Bigr]
	\end{aligned}
\end{equation}
where we have redefined the function $C_i$. Due to the independence of the tensorial structures listed in the equation above, now all the coefficients $C_i$ need to vanish
\begin{equation}
	C_i=0.
\end{equation}
In particular, $C_1=0$ and $C_2=0$ are differential equation of the second order, called primary equations. Their explicit form is given by 
\begin{equation}\label{eq:primeqscwi} 
	\begin{aligned}
		&K_{31}A=0,\\&
		K_{32}A=0
	\end{aligned}
\end{equation}
where we have introduced the operator \cite{Bzowski:2013sza}
\begin{equation}
	K_i=\frac{\partial^2}{\partial p_i^2}+\frac{\left(d+1-2 \Delta_i\right)}{p_i} \frac{\partial}{\partial p_i}, \qquad\qquad K_{i j}=K_i-K_j. \\
\end{equation}
acting on the single form factor $A(p_1,p_2,p_3)$ of the transverse traceless sector. \\
The operators $K_{ij}$ allow to relate the differential content of the equations for the form factors, which are combinations of Appell hypergeometric functions \cite{Coriano:2013jba,Coriano:2018bbe} of two variables, briefly discussed in Appendix \ref{appendix:3kint}. \\
The remaining special CWIs ($C_i=0$ with $i=\{3,4,5\}$) are differential equations of the first order, i.e. the secondary equations. Their explicit expressions are given by 
\begin{equation} \label{eq:seceqscwi} 
	\begin{aligned}
		&0=A-p_1\frac{\partial A }{\partial p_1},\\
		&0=\frac{\partial A }{\partial p_2},\\
		&0=
		-2 \, \frac{p_1^2-2p_2^2+2p_3^2}{p_1^2}A
		-\frac{p_1^2+p_2^2-p_3^2}{p_1}\frac{\partial A}{\partial p_1}-4p_2\frac{\partial A}{\partial p_2}.
	\end{aligned}
\end{equation}

\subsection{Solving the CWIs}
The solution of the conformal Ward identities for the $\langle TJO\rangle$ can be written in terms of integrals involving a product of three Bessel functions, namely 3K integrals  \cite{Bzowski:2013sza,Bzowski:2015yxv}, as illustrated in Appendix \ref{appendix:3kint}. 4K extensions of these formulations appear in the large energy $s$ and momentum transfer $t$ limits of four-point functions in CFT, as special combinations of Lauricella functions \cite{Coriano:2019nkw}. 
In the 3K case, we recall the definition
\begin{equation}
	I_{\alpha\left\{\beta_1, \beta_2, \beta_3\right\}}\left(p_1, p_2, p_3\right) \equiv  \int d x x^\alpha \prod_{j=1}^3 p_j^{\beta_j} K_{\beta_j}\left(p_j x\right)
\end{equation}
where $K_\nu$ is the modified Bessel function of the second kind 
\begin{equation}
	K_\nu(x)=\frac{\pi}{2} \frac{I_{-\nu}(x)-I_\nu(x)}{\sin (\nu \pi)}, \qquad \nu \notin \mathbb{Z} \qquad\qquad I_\nu(x)=\left(\frac{x}{2}\right)^\nu \sum_{k=0}^{\infty} \frac{1}{\Gamma(k+1) \Gamma(\nu+1+k)}\left(\frac{x}{2}\right)^{2 k}
\end{equation}
with the property
\begin{equation}
	K_n(x)=\lim _{\epsilon \rightarrow 0} K_{n+\epsilon}(x), \quad n \in \mathbb{Z}.
\end{equation}
We will also use a reduced version of the 3K integral defined as
\begin{equation}
	J_{N\left\{k_1,k_2,k_3\right\}}=I_{\frac{d}{2}-1+N\left\{\Delta_j-\frac{d}{2}+k_j\right\}}.
\end{equation}
The 3K integrals satisfy a relation analogous to the dilatation equation with scaling degree  \cite{Bzowski:2013sza}
\begin{equation}
	\text{deg}\left(J_{N\left\{k_1,k_2,k_3\right\}}\right)=\Delta_t+k_t-2 d-N,
\end{equation}
where 
\begin{equation}
	k_t=k_1+k_2+k_3,\qquad\qquad \Delta_t=\Delta_1+\Delta_2+\Delta_3.
\end{equation}
From this analysis, it is straightforward to relate the form factor $A$ to the 3K integrals. Indeed, the dilatation Ward identity \eqref{eq:dilatAconst} tells us that the form factor $A$ can be written as a combination of integrals of the following type
\begin{equation}
	J_{3+k_t,\{k_1,k_2,k_3\}}.
\end{equation}
The special CWIs fix the remaining indices $k_1$, $k_2$ and $k_3$.
Recalling the following property of the 3K integrals
\begin{equation}
	K_{n m} J_{N\left\{k_j\right\}}=-2 k_n J_{N+1\left\{k_j-\delta_{j n}\right\}}+2 k_m J_{N+1\left\{k_j-\delta_{j m}\right\}},
\end{equation}
we can write the most general solution of the primary equations \eqref{eq:primeqscwi} as
\begin{equation}\label{eq:TJOsolprimu}
	A=c_1 J_{3\{0,0,0\}}\equiv c_1 I_{4\{2,1,\Delta_3-2\}},
\end{equation} 
where $c_1$ is an arbitrary constant. When dealing with such a 3K integral, one needs to be careful as 
$J_{3\{0,0,0\}}$ may diverge. Depending on the value of $\Delta_3$, a regularization may be necessary. 
In general, it can be shown that the 3K integral $I_{\alpha\{\beta_1,\beta_2,\beta_3\}}$ diverges if
\begin{equation}
	\alpha+1 \pm \beta_1 \pm \beta_2 \pm \beta_3=-2 k \quad, \quad k=0,1,2, \dots
\end{equation}
For a more detailed review of the topic, see Appendix \ref{appendix:3kint} and \cite{Bzowski:2013sza,Bzowski:2015pba,Bzowski:2015yxv}.
If the above condition is satisfied, we need to regularize the integral. This can be done by
shifting the parameters of the 3K integrals
\begin{equation}
	I_{\alpha\left\{\beta_1, \beta_2, \beta_3\right\}} \rightarrow I_{\alpha+u \epsilon\left\{
		\beta_1+v_1 \epsilon, \beta_2+v_2 \epsilon, \beta_3+v_3 \epsilon
		\right\}}, \qquad \quad J_{N\left\{k_1, k_2, k_3\right\}} \rightarrow J_{N+u \epsilon\left\{k_1+v_1 \epsilon, k_2+v_2 \epsilon, k_3+v_3 \epsilon\right\}}
\end{equation}
or equivalently
\begin{equation}
	d \rightarrow 4+2 u \epsilon \quad  \quad \Delta_i \rightarrow \Delta_i+\left(u+v_i\right) \epsilon .
\end{equation}
In general, the regularization parameters $u$ and $v_i$ are arbitrary, though in certain cases there can be specific constraints on them. \\
In any case, regardless of the implementation of a regularization, after inserting our solution \eqref{eq:TJOsolprimu} back into the secondary equations \eqref{eq:seceqscwi}, one finds $c_1=0$. Therefore,  we have shown that the conformal constraints require the parity-odd sector of the correlator to be zero
\begin{equation}
	\left\langle T^{ \mu_1\nu_1}\left(p_1\right) 
	J^{   \mu_2}\left(p_2\right) O	\left(p_3\right)\right\rangle_{odd} =0.
\end{equation}
Considering the fact that the parity-even sector of the $\langle TJO\rangle$ vanishes as well \cite{Bzowski:2013sza}, we come to the conclusion that conformal symmetry prohibits the off-shell interaction of a graviton with a photon and a scalar/pseudoscalar, such as a dilaton/axion. In other words, a gravitational field cannot induce an axion to a spin-1 transition in the presence of conformal symmetry.

\section{The conformal $\langle JJO\rangle_{odd}$ and $\langle TTO \rangle_{odd}$\, : a reappraisal}
\label{sec4}
Let us now discuss the $\langle JJO\rangle_{odd}$ and $\langle TTO\rangle_{odd}$ correlators, which have been investigated in \cite{Coriano:2023cvf}. The procedure follows the same approach of the $\langle TJO\rangle_{odd}$ discussed in the previous section.\\
First, we will briefly review the analysis of the $\langle JJO\rangle_{odd}$ correlator.
Then, we will proceed with a re-evaluation of the $\langle TTO\rangle_{odd}$, as the primary and secondary equations in the published version of \cite{Coriano:2023cvf} were not correctly identified. Here, we present the correct version of these equations and solve them accordingly. The previous solution reported in \cite{Coriano:2023cvf} for the $\langle TTO\rangle_{odd}$ with $\Delta_3 = 4$ remains valid, and the conclusions and interpretations of that paper are unaffected. However, these corrections have led us to discover new non-vanishing solutions for the $\langle TTO\rangle_{odd}$ characterized by different values of $\Delta_3$.
Lastly, we further elaborate on all the solutions of both the $\langle JJO\rangle_{odd}$ and $\langle TTO\rangle_{odd}$, offering a new interpretation using the shadow transform.

\subsection{The $\langle JJO\rangle_{odd}$ }
We start by considering the following conservation Ward identity
\begin{equation}\label{eq:jjoconservedid}
	\begin{aligned}
		p_{i\mu_i} \langle J^{\mu_1}(p_1)J^{\mu_2}(p_2)O(p_3)\rangle_{odd}=0\qquad \qquad i=1,2.
	\end{aligned}
\end{equation}
Such equation is satisfied independently of the fact that $J$'s are vector or axial-vector currents.
Indeed, the chiral anomaly does not affect the $\langle JJO\rangle$ correlator.
Due to this equation, the correlator comprises only a transverse part, which can be formally expressed by introducing a single form factor $A(p_1, p_2, p_3)$
\begin{equation}
	\left\langle J^{ \mu_1}\left(p_1\right) 
	J^{   \mu_2}\left(p_2\right) O	\left(p_3\right)\right\rangle_{odd}=\pi^{\mu_1}_{\alpha_1}\left(p_1\right)
	\pi^{\mu_2}_{\alpha_2} \left(p_2\right) \Bigl[  A(p_1,p_2,p_3)\varepsilon^{\alpha_1\alpha_2 p_1 p_2} \Bigr]=A(p_1,p_2,p_3)\varepsilon^{\alpha_1\alpha_2 p_1 p_2}.
\end{equation}
Note that in this case the projectors $\pi^{\mu_i}_{\alpha_i}\left(p_i\right)$ can be omitted since they act as an identity on the tensorial structure in the brackets.\\
We now examine how the conformal constraints fix the form factor $A(p_1,p_2,p_3)$.
The invariance of the correlator under dilatation is reflected in the equation
\begin{align}\label{eq:diljjouu}
	\sum_{i=1}^{3} p_i \frac{\partial A}{\partial p_i }-\left(\sum_{i=1}^3\Delta_i-2d- 2\right) A=0.
\end{align}
The special conformal constraints are instead given by the following primary equations 
\begin{equation}
	\begin{aligned} \label{eq:primOmog} 
		K_{31}A=0,\\
		K_{32}A=0,
	\end{aligned}
\end{equation}
and the secondary equations 
\begin{equation}\label{eq:secondaryJJOuu}
	\begin{aligned}
		0=p_2\frac{\partial A }{\partial p_2}+ (d-1-\Delta_2)A,\\
		0=p_1\frac{\partial A }{\partial p_1}+ (d-1-\Delta_1)A.
	\end{aligned}
\end{equation}
We can write the most general solution of the dilatations equation \eqref{eq:diljjouu} and the primary equations \eqref{eq:primOmog} as
\begin{equation}\label{eq:solprimarieJJO}
	A=c_1 J_{2\{0,0,0\}}\equiv c_1 I_{3\{1,1,\Delta_3-2\}},
\end{equation} 
where $c_1$ is an arbitrary constant.
When dealing with such a 3K integral, one needs to be careful as 
$J_{2\{0,0,0\}}$ may diverge (see Appendix \ref{appendix:3kint}). Depending on the value of $\Delta_3$, a regularization may be necessary. Then, by inserting the (regularized) solution \eqref{eq:solprimarieJJO} back into the secondary equations \eqref{eq:secondaryJJOuu}, one is able to determine if $c_1=0$, depending on the value of $\Delta_3$. This procedure has been carried out in \cite{Coriano:2023cvf}.\\
After explicitly expressing the 3K integral,
the general conformal $\langle JJO\rangle_{odd}$ correlator can be written as
\begin{equation}\label{eq:jjosol}
	\begin{aligned}
		& \left\langle J^{\mu_1}\left(p_1\right) J^{\mu_2}\left(p_2\right) O_{\left(\Delta_3 \neq 0,4\right)}\left(p_3\right)\right\rangle_{odd}=0, \\
		& \left\langle J^{\mu_1}\left(p_1\right) J^{\mu_2}\left(p_2\right) O_{\left(\Delta_3=0\right)}\left(p_3\right)\right\rangle_{odd}=\frac{c_1}{p_3^4} \varepsilon^{p_1 p_2 \mu_1 \mu_2},\\
		& \left\langle J^{\mu_1}\left(p_1\right) J^{\mu_2}\left(p_2\right) O_{\left(\Delta_3=4\right)}\left(p_3\right)\right\rangle_{odd}=c_1 \varepsilon^{p_1 p_2 \mu_1 \mu_2} \\
	\end{aligned}
\end{equation}
Notice the presence of a double pole in the $\Delta_3=0$ case. 

\subsection{The $\langle TTO\rangle_{odd}$ }
For this correlator, we start by considering the conservation and trace Ward identities for the energy-momentum tensor
\begin{equation} \label{eq:conswittoa}
	p_{i\mu_i} \left\langle T^{ \mu_1\nu_1}\left(p_1\right) 
	\right.\left.
	T^{ \mu_2\nu_2}\left(p_2\right) O	\left(p_3\right)\right\rangle_{odd} =0, \qquad\qquad 
	g_{\mu_i\nu_i} \left\langle T^{ \mu_1\nu_1}\left(p_1\right) 
	\right.\left.
	T^{ \mu_2\nu_2}\left(p_2\right) O	\left(p_3\right)\right\rangle_{odd} =0,
\end{equation}
for $i=\{1,2\}$. 
Due to such equations, the correlator comprises only a transverse-traceless part which can be formally expressed in terms of two form factors
\begin{equation}
	\begin{aligned}
		\left\langle T^{ \mu_1\nu_1}\left(p_1\right) 
		\right.&\left.
		T^{ \mu_2\nu_2}\left(p_2\right) O	\left(p_3\right)\right\rangle_{odd} =
		\\
		&\Pi^{\mu_1\nu_1}_{\alpha_1\beta_1}\left(p_1\right)
		\Pi^{\mu_2\nu_2}_{\alpha_2\beta_2} \left(p_2\right) \Bigl[  A_1(p_1,p_2,p_3)\varepsilon^{\alpha_1\alpha_2 p_1 p_2}p^{\beta_1}_2p_3^{\beta_2}
		+A_2(p_1,p_2,p_3)\varepsilon^{\alpha_1\alpha_2 p_1 p_2}\delta^{\beta_1\beta_2} 
		\Bigr].
	\end{aligned}
\end{equation}
The invariance of the $\langle TTO \rangle$ under dilatation is reflected in the following constraints on the form factors
\begin{equation}
	\begin{aligned}
		\sum_{i=1}^{3} p_i \frac{\partial A_1}{\partial p_i }-\left(\sum_{i=1}^3\Delta_i-2d- 4\right) A_1=0,\\
		\sum_{i=1}^{3} p_i \frac{\partial A_2}{\partial p_i }-\left(\sum_{i=1}^3\Delta_i-2d- 2\right) A_2=0.
	\end{aligned}
\end{equation}
The special conformal constraints are  instead given by the following primary equations\footnote{These equations rectify those that were incorrectly identified in \cite{Coriano:2023cvf}.} 
\begin{equation}\label{eq:primaryTTOn}
	\begin{aligned}
		&K_{31}A_1=0,\qquad 
		&&K_{32}A_1=0,\\
		&K_{31}A_2-2p_2 \frac{\partial A_1}{\partial p_2 }+4 A_1=0, \qquad
		&&K_{32}A_2-2p_1 \frac{\partial A_1}{\partial p_1 }+4 A_1=0.
	\end{aligned}
\end{equation}
and the secondary equations
\begin{equation} \label{eq:seceqtto}
	\begin{aligned}
		&0=(p_1^2-p_2^2-p_3^2)A_1+2A_2+2p_1p_2^2\frac{\partial A_1}{\partial p_1}-p_2(p_1^2+p_2^2-p_3^2)\frac{\partial A_1}{\partial p_2}-2p_2\frac{\partial A_2}{\partial p_2}, \\
		&0=(p_1^2-p_2^2+p_3^2)A_1-2A_2-2p_1^2p_2\frac{\partial A_1}{\partial p_2}+p_1(p_1^2+p_2^2-p_3^2)\frac{\partial A_1}{\partial p_1}+2p_1\frac{\partial A_2}{\partial p_1}, \\
		&0=2\left(  \frac{p_1^2+2p_2^2-2p_3^2}{p_1^2}  \right) A_1+\frac{8}{p_1^2}A_2 -\frac{p_1^2+p_2^2-p_3^2}{p_1}
		\frac{\partial A_1}{\partial p_1}-\frac{2}{p_1}\frac{\partial A_2}{\partial p_1}-4p_2\frac{\partial A_1}{\partial p_2},\\
		&0=-2\left( \frac{2p_1^2+p_2^2-2p_3^2}{p_2^2}\right)A_1-\frac{8}{p_2^2}A_2+
		4p_1\frac{\partial A_1}{\partial p_1}+\frac{p_1^2+p_2^2-p_3^2}{p_2}\frac{\partial A_1}{\partial p_2}+\frac{2}{p_2}\frac{\partial A_2}{\partial p_2}.
	\end{aligned}
\end{equation}
In order to solve the primary eqs$.$ \eqref{eq:primaryTTOn}, we rewrite them as a set of homogeneous equations by repeatedly applying the
operator $K_{ij}$ on them
\begin{equation}
	\begin{aligned}
		&K_{31}A_1=0,\qquad 
		&&K_{32}A_1=0,\\
		&K_{31}K_{31}A_2=0, \qquad
		&&K_{32}K_{32}A_2=0.
	\end{aligned}
\end{equation}
The most general solution to these equations can then be written in terms of the following combinations of 3K integrals
\begin{equation}
	\begin{aligned} \label{eq:solomogtto}
		&A_1=c_1 J_{4\{0,0,0\}},\\
		&A_2=c_2 J_{3\{1,0,0\}}+c_3 J_{3\{0,1,0\}}+c_4 J_{3\{0,0,1\}}+c_5 J_{4\{1,1,0\}}+c_6J_{2\{0,0,0\}}.
	\end{aligned}
\end{equation} 
We then insert these solutions back into the non-homogeneous primary eqs$.$ \eqref{eq:primaryTTOn} and the secondary eqs$.$ \eqref{eq:seceqtto} in order to fix the constants $c_i$. We can solve such constraints for different values of the conformal dimensions $\Delta_3$.
Such procedure involves a possible regularization,  the use of the properties of 3K integrals and their limits $p_i\rightarrow0$ described in the Appendix \ref{appendix:3kint}. For odd values of $\Delta_3$, no regularization is needed, and we found only vanishing solutions. We also considered different examples with even values of $\Delta_3$. In particular, we found

\begin{allowdisplaybreaks}
\begin{align}\label{eq:ttosol}
	&
	\left\langle T^{ \mu_1\nu_1}\left(p_1\right) 
	T^{ \mu_2\nu_2}\left(p_2\right) O_{\left(\Delta_3=-2 \right)}	\left(p_3\right)\right\rangle_{odd} =
	\Pi^{\mu_1\nu_1}_{\alpha_1\beta_1}\left(p_1\right)
	\Pi^{\mu_2\nu_2}_{\alpha_2\beta_2} \left(p_2\right)  \,  \times\nonumber \\ &\qquad \frac{c_1}{p_3^8}\Biggl[  
	\left(3\left(p_1^2-p_2^2\right)^2-2\left(p_1^2+p_2^2\right)p_3^2-p_3^4\right)
	\varepsilon^{\alpha_1\alpha_2 p_1 p_2}\delta^{\beta_1\beta_2} 
	-2\left(3p_1^2+3p_2^2+p_3^2\right)\varepsilon^{\alpha_1\alpha_2 p_1 p_2}p^{\beta_1}_2p_3^{\beta_2}
	\Biggr],\nonumber \\[8pt]
	&
	\left\langle T^{ \mu_1\nu_1}\left(p_1\right) 
	T^{ \mu_2\nu_2}\left(p_2\right) O_{\left(\Delta_3=0 \right)}	\left(p_3\right)\right\rangle_{odd} =
	\nonumber\\ &\qquad\qquad
	\Pi^{\mu_1\nu_1}_{\alpha_1\beta_1}\left(p_1\right)
	\Pi^{\mu_2\nu_2}_{\alpha_2\beta_2} \left(p_2\right) \, \frac{c_1}{p_3^4}\Biggl[  
	-\frac{p_1^2+p_2^2-p_3^2}{2}
	\varepsilon^{\alpha_1\alpha_2 p_1 p_2}\delta^{\beta_1\beta_2} 
	+\varepsilon^{\alpha_1\alpha_2 p_1 p_2}p^{\beta_1}_2p_3^{\beta_2}
	\Biggr],\nonumber\\[8pt]
	&
	\left\langle T^{ \mu_1\nu_1}\left(p_1\right) 
	T^{ \mu_2\nu_2}\left(p_2\right) O_{\left(\Delta_3=2 \right)}	\left(p_3\right)\right\rangle_{odd} =0
	,\nonumber\\[8pt] 
	&
	\left\langle T^{ \mu_1\nu_1}\left(p_1\right) 
	T^{ \mu_2\nu_2}\left(p_2\right) O_{\left(\Delta_3=4 \right)}	\left(p_3\right)\right\rangle_{odd} =\nonumber
	\\ &\qquad\qquad
	\Pi^{\mu_1\nu_1}_{\alpha_1\beta_1}\left(p_1\right)
	\Pi^{\mu_2\nu_2}_{\alpha_2\beta_2} \left(p_2\right) \, c_1\Biggl[  
	-\frac{p_1^2+p_2^2-p_3^2}{2}
	\varepsilon^{\alpha_1\alpha_2 p_1 p_2}\delta^{\beta_1\beta_2} 
	+\varepsilon^{\alpha_1\alpha_2 p_1 p_2}p^{\beta_1}_2p_3^{\beta_2}
	\Biggr],\nonumber\\[8pt]
	&
	\left\langle T^{ \mu_1\nu_1}\left(p_1\right) 
	T^{ \mu_2\nu_2}\left(p_2\right) O_{\left(\Delta_3=6 \right)}	\left(p_3\right)\right\rangle_{odd} =
	\Pi^{\mu_1\nu_1}_{\alpha_1\beta_1}\left(p_1\right)
	\Pi^{\mu_2\nu_2}_{\alpha_2\beta_2} \left(p_2\right)  \,  \times \nonumber\\ &\qquad c_1\Biggl[  
	\left(3\left(p_1^2-p_2^2\right)^2-2\left(p_1^2+p_2^2\right)p_3^2-p_3^4\right)
	\varepsilon^{\alpha_1\alpha_2 p_1 p_2}\delta^{\beta_1\beta_2} 
	-2\left(3p_1^2+3p_2^2+p_3^2\right)\varepsilon^{\alpha_1\alpha_2 p_1 p_2}p^{\beta_1}_2p_3^{\beta_2}
	\Biggr],\nonumber\\[8pt]
	&
	\left\langle T^{ \mu_1\nu_1}\left(p_1\right) 
	T^{ \mu_2\nu_2}\left(p_2\right) O_{\left(\Delta_3=8 \right)}	\left(p_3\right)\right\rangle_{odd} =0.
\end{align}
\end{allowdisplaybreaks}

\subsection{Connections with the anomalies}
As we have seen in the previous sections, specifically in eq$.$ \eqref{eq:jjosol} and \eqref{eq:ttosol}, the  $\langle JJO\rangle_{odd}$ and $\langle TTO \rangle_{odd}$  do not always vanish. Indeed, such correlators are protected by chiral and conformal anomalies. 
Specifically, we can justify the existence of the non-vanishing solutions with $\Delta_3 = 4$ by providing an example of a scalar operator $O$ which is linked to the chiral anomaly
\begin{equation}
	O = \nabla_\mu J_A^\mu=  \mathcal{A}_{chiral}=a_1 \, \varepsilon^{\mu \nu \rho \sigma} F_{\mu \nu} F_{\rho \sigma}+a_2 \, \varepsilon^{\mu \nu \rho \sigma} R_{\beta \mu \nu}^\alpha R_{\alpha \rho \sigma}^\beta.
\end{equation}
Choosing such scalar operator, then we can  write
\begin{equation}
	\begin{aligned}
		&\langle J^{\mu_1}(x_1)J^{\mu_2}(x_2)\, O(x_3)\rangle_{odd} =\frac{\delta  \mathcal{A}_{chiral}(x_3)}{\delta A_{\mu_1}(x_1)\delta A_{\mu_2}(x_2)}\bigg|_{g_{\mu\nu}=\eta_{\mu\nu},\, A_\mu=0},\\&
	\langle T^{\mu_1\nu_1}(x_1)T^{\mu_2\nu_2}(x_2)\,O(x_3)\rangle_{odd}=4 \frac{\delta  \mathcal{A}_{chiral}(x_3)}{\delta g_{\mu_1\nu_1}(x_1)\delta g_{\mu_2\nu_2}(x_2)}\bigg|_{g_{\mu\nu}=\eta_{\mu\nu},\, A_\mu=0}
	\end{aligned}
\end{equation}
in the flat limit.
By Fourier transforming these expressions, one obtains exactly the solutions in eq$.$ \eqref{eq:jjosol} and \eqref{eq:ttosol} with $\Delta_3 = 4$.
Note that $O=\nabla \cdot J_A$ is a primary operator since, by acting on it with the special conformal operator $\mathcal{K}$, we obtain a vanishing result
\begin{equation}
	\mathcal{K}_\nu P_\mu\left|J_A^\mu\right\rangle  =\left[\mathcal{K}_\nu, P_\mu\right]\left|J_A^\mu\right\rangle =2\left(\mathcal{D}\,  \delta_{\mu \nu}-M_{\nu \mu}\right)\left|J_A^\mu\right\rangle  =2(\Delta-d+1)\left|J_{A\, \nu}\right\rangle =0
\end{equation}
where $\mathcal{D}$ is the dilatation operator and in the last passage we used the fact that the conformal dimension of $J_A$ is $\Delta=d-1$.\\
Besides $O=\nabla \cdot J_A$, one could alternatively consider $O=g_{\mu\nu}T^{\mu\nu}$ as a scalar operator with $\Delta_3 = 4$. The correlators would then be similarly protected by the possible existence  of parity-odd trace anomalies in CFT
\begin{equation}
		g_{\mu \nu}\left\langle T^{\mu \nu}\right\rangle_{odd}=f_1 \varepsilon^{\mu \nu \rho \sigma} R_{\alpha \beta \mu \nu} R_{\rho \sigma}^{\alpha \beta}+f_2 \varepsilon^{\mu \nu \rho \sigma} F_{\mu \nu} F_{\rho \sigma} .
\end{equation}
 This last hypothesis has been analyzed in depth in \cite{Coriano:2023cvf}.

\subsection{The shadow transforms and the $\Delta_3\leq 0$ solutions}
In the previous section, we justified the non-vanishing solutions of the $\langle JJO\rangle_{odd}$ and $\langle TTO\rangle_{odd}$ correlators with $\Delta_3=4$ through their connection to the chiral and conformal anomalies. However, these are not the only non-vanishing solutions that we have discussed. Indeed, such correlators can also be non-zero when $\Delta_3=0$. \\
The general solution of these correlators with $\Delta_3=4$ and $\Delta_3=0$ differ by a factor $p_3^4$ .
Although the condition $\Delta_3=0$ is non-physical and violate unitarity, we can now ask ourselves if there are any justifications for the existence of such non-vanishing solutions in CFT. As we will see, the conformal and chiral anomalies can still be responsible for such solutions. To demonstrate this, we need to introduce the concept of shadow transform, which we will briefly review \cite{Ferrara:1972uq,Simmons-Duffin:2012juh,Anninos:2017eib}.\\
Given a primary operator $O_{l,\Delta}$ of spin $l$ and scaling dimension $\Delta$, we can construct a {\em shadow} primary field in $d$ dimensions in the following way
\begin{equation}
	\tilde{O}_{l, \bar{\Delta}}(x)=\int d^d y\, G_{l, \bar{\Delta}}(x-y) O_{l, \Delta}(y)
\end{equation}
with spin $l$ and conjugate scaling dimension 
\begin{equation}
	\bar{\Delta}=d-\Delta.
\end{equation}
The kernel $G_{l,\bar{\Delta}}(x-y)$ takes the form of a 2-point function, with spin $l$ and dimension $\bar{\Delta}$ operators, in a $d$-dimensional CFT. The integral defines the shadow transform. In particular, for a scalar operator we have
\begin{equation}\label{eq:stscalar}
	\tilde{O}_{\bar{\Delta}}(x)=\int d^d y \frac{c_{\bar{\Delta}}}{|x-y|^{2 \bar{\Delta}}} O_{{\Delta}}(y).
\end{equation}
The constant $c_{\bar{\Delta}}$ is a normalization factor, which we leave arbitrary for now.
What is special about this particular choice of integration kernel - as opposed to say a kernel of the form $|x-y|^{-2\Delta'}$ for some generic $\Delta'$ - is that the resulting object $\tilde{O}(x)$ transforms again as a local primary field under the conformal group.\\
 The inverse of a shadow transform is again a shadow transform. For example, the scalar relation \eqref{eq:stscalar} may be inverted as 
\begin{equation}
	O_{{\Delta}}(x)=\int d^d y \frac{\tilde{c}_{{\Delta}}}{|x-y|^{2 \Delta}} \tilde{O}_{\bar{\Delta}}(y).
\end{equation}
This becomes obvious in the momentum space description. The Fourier transform of the expression \eqref{eq:stscalar} can be computed by using the following relation
\begin{equation}
	\int d^d x \, x^{-2 \bar{\Delta}} e^{-i k \cdot x}=\frac{\pi^{d / 2}\, 2^{d-2 \bar{\Delta}}\, \Gamma\left[\frac{d-2 \bar{\Delta}}{2}\right]}{\Gamma[\bar{\Delta}]} k^{2 \bar{\Delta}-d} .
\end{equation}
In particular, for a suitable choice of $c_{\bar{\Delta}} $ we end up with
\begin{equation}
	\tilde{O}_{\bar{\Delta}}(k)=k^{2 \bar{\Delta}-d} O_\Delta(k).
\end{equation}
Given this definition, we now re-examine the solutions of the $\langle JJO\rangle_{odd}$ and $\langle TTO\rangle_{odd}$ in eq$.$ \eqref{eq:jjosol} and \eqref{eq:ttosol}. 
For both correlators, the solution with $\Delta_3= 0$ is connected to the one with $\Delta_3 = 4$ through a shadow transformation of the scalar operator $O$.\\
In the previous sections, we have provided examples of non-vanishing $\langle JJO\rangle_{odd}$ and $\langle TTO\rangle_{odd}$ correlators, using $\nabla \cdot J_A$ or $g_{\mu\nu} T^{\mu\nu}$ as scalar operators with $\Delta_3 = 4$. Applying a shadow transform to these specific scalar operators yields non-vanishing correlators that satisfy eqs. \eqref{eq:jjosol} and \eqref{eq:ttosol} for $\Delta_3 = 0$. This demonstrates that the chiral and conformal anomaly content of the quantum expectation values of these operators, can explain the non-vanishing nature of these other solutions too.\\
To conclude, it is worth noting that for the $\langle TTO\rangle_{odd}$ correlator, we have identified additional non-zero solutions, specifically for $\Delta_3 = -2$ and 6. These solutions exhibit a more complex structure compared to those with $\Delta_3 = 0$ and 4, and they do not seem to be directly related to the functional derivatives of the anomalies that we have previously discussed. It would be interesting to find a justification for the existence of these solutions as well. Notably, the solutions with $\Delta_3 = -2$ and 6 are also related to each other by a shadow transform.

\section{Physical implications for chiral/gravitational backgrounds}
\label{sec5}
Our analysis has been set to explore the possible implication of conformal symmetry of 3-point functions whenever scalar operators appear in mixed correlators which have a direct physical relevance. 
Obviously, the use of conformal symmetry limits the generality of the result, for being surely specific, and it is then natural to ask under which conditions these results can be applied. \\
One possibility, but surely not the only one, is encountered in early universe cosmology, when conformal symmetry is expected to have played a significant role before that any physical scale appeared in the dynamics. \\
Indeed, scalar and pseudoscalar fields are considered potential components of dark matter in current cosmological models, although definitive evidence for their existence has yet to be collected. 
Other important applications are natural inflation models \cite{Adams:1992bn}, which involve a pseudoscalar field (the inflaton) with a periodic potential inspired by axion-like particles.  Mechanisms for baryogenesis also involve scalar fields carrying baryon or lepton number. If these fields are pseudoscalars, they can exhibit CP-violating interactions that generate a baryon asymmetry \cite{Affleck:1984fy}.\\
Pseudoscalar fields, in general, can generate distinctive signatures if they couple to gravity, detectable through their imprints on the stochastic background of gravitational waves. In the case of axion-like fields and their coupling to gauge fields, they can induce spin-1 helicity asymmetries before any spontaneously broken phase intervenes. In this context, a correlator such as the $\langle TTO\rangle $ for example, where $ O$  is a pseudoscalar coupled to 
an axion-like field $\phi$, is directly connected with a gravitational anomaly. The local effective action describing this interaction in the infrared, once a conformal symmetry breaking scale $(\Lambda)$ is present, is of the form  
\beq
\label{sub}
\mathcal{L}_{axion} \supset \frac{\phi}{\Lambda} \varepsilon^{\mu \nu \rho \sigma} R_{\beta \mu \nu}^\alpha R_{\alpha \rho \sigma}^\beta
\eeq
This correlator mediates an interaction between the pseudoscalar and two gravitational waves at semiclassical level and is part of the effective action generated by integrating out conformal matter in quantum corrections. The interaction is the analogue of the $(\phi/f) F\tilde{F}$ interaction of ordinary axion-physics, which is associated with the $\langle JJO\rangle $ $(\Delta_3=4)$ in \eqref{eq:jjosol}.
This process, referred to as {\em conformal backreaction} \cite{Coriano:2022ftl}, is based on the hypothesis that the universe underwent a conformal phase before any ordinary mechanism based on spontaneous symmetry breaking took place.\\
 This assumption is supported by the fact that ordinary gauge theories in their exact phases are classically scale-invariant and manifest conformal and chiral anomalies in their fermion sectors.  Both the $\langle TTO\rangle $ and the $\langle JJO\rangle$  correlators are part of the gravitational effective action, where semiclassical corrections modify the interaction of gravity with other fields and are essential for investigating the chiral behavior in the spectrum of gravitational waves. \\
The $\langle JJO\rangle $, where $J$  is a spin-1 current to gauge fields like the hypercharge 
$ Y $ and axion-like fields, can also be linked to gravity non-perturbatively. 
In this case, $O $ interpolates with a trace anomaly of a stress-energy tensor of odd parity. This point has come to the attention of several groups in the recent literature. The appearance of this anomaly with Standard Model fermions has been attributed to the implementation of regularization procedures in the perturbative analysis of such contributions  with conflicting results \cite{Armillis:2010pa,Abdallah:2023cdw,Abdallah:2021eii,Bonora:2017gzz,Bonora:2014qla,Bastianelli:2019zrq,Larue:2023tmu,Bonora:2018obr}.\\
In a previous analysis \cite{Coriano:2023cvf}, we showed that the CWIs permit a minimal form of this interaction. However, such approach is non-perturbative. In this context, the axion/dilaton field $(\chi)$ would interact according to the form described in \eqref{sub}.\\
Another possibility is to consider $O\sim \partial\cdot J_{CS}$, representing the divergence of a Chern-Simons (CS) current. This idea was discussed in \cite{Dolgov:1987yp}, where the current generates an anomaly for a spin-1 particle \cite{delRio:2020cmv,delRio:2021bnl,Agullo:2018nfv}. This anomaly is associated with a CS current of the form 
\beq
J_{CS}^{\lambda} =\epsilon^{\lambda \mu\nu\rho} V_{\mu}\partial_{\nu} V_{\rho}.
\eeq
The $J_{CS}$ current mediates the gravitational chiral anomaly with spin-1 virtual particles in the loops, creating an asymmetry between their two circular modes and inducing optical helicity \cite{Galaverni:2020xrq}.
In general, a chiral chemical potential can generate chiral currents with longitudinal components to which pseudoscalar fields can couple (see also \cite{Kamada:2022nyt} for a general discussion of chirality effects in astrophysics). The amplitude is anomaly-mediated and the conformal symmetry of the background is, in this case, broken. However, it has been shown \cite{Coriano:2024nhv} that finite density corrections do not affect the chiral anomaly form factor identified from the covariant expansion of the anomaly vertex, which remains independent of both the chemical potential and temperature corrections \cite{Coriano:2024rcd}. This implies that the propagation of the chiral fermions in the loop is essentially conformal, and is described  by the exchange of a massless anomaly pole. Corrections related to the chemical potential are proportional to the fermion mass terms, but leave the anomaly contribution as in the conformal limit \cite{Coriano:2023hts}. \\
A final comment concerns the $\langle TJO\rangle $ that vanished identically as an off-shell correlator in a CFT, both in its even and odd components. This result can be interpreted as a constraint on the absence of an axion/dilaton to a spin-1 mixing in the presence of an external gravitational field. This can be viewed as a constraint similar to the axion to photon transition in the presence of an external magnetic field, which is the basic process for axion detection with helioscopes \cite{Zioutas:2010zz}. 

\section{Conclusions}
We have analyzed all possible 3-point functions constructed with the energy-momentum tensor, currents and at least one scalar operator in CFT. Most of the correlators are constrained to be zero, except some specific cases which are protected from vanishing by their anomalies.\\
In particular, all the correlators built with at least two scalar operators vanish because of Lorentz invariance. There is no need to require invariance under the full conformal group, as it is not even possible to construct tensorial structures for them.\\
The analysis of the $\langle TJO \rangle_{odd}$ is more intricate and requires the full conformal group in order to determine the correlator. Specifically, we have shown that such correlator vanishes in a CFT. The even part of $\langle TJO \rangle_{odd}$ has been investigated in \cite{Bzowski:2013sza} and it was found to be zero as well.\\
Among all the correlators we have examined, the only ones not constrained to vanish are the $\langle JJO\rangle_{odd} $ and $\langle TTO\rangle_{odd} $. Interestingly, this effect is not visible in coordinate space, as these correlators were found to be zero in that context \cite{Jain:2021wyn}.
However, in momentum space, such correlators cannot always vanish since they are protected by the chiral and conformal anomalies when $\Delta_3=0$ and 4.
For instance, one can consider as scalar operators $O=\nabla \cdot J_A$ or $O=g_{\mu\nu}T^{\mu\nu}$ and therefore the chiral/conformal anomaly prevent the correlators from vanishing when $\Delta_3=4$.
Furthermore, the non-zero solutions we found for the $\langle JJO\rangle_{odd}$ and $\langle TTO\rangle_{odd}$ with $\Delta_3 = 0$ can be obtained by selecting the shadow transform of $\nabla \cdot J_A$ or $g_{\mu\nu} T^{\mu\nu}$ as scalar operator $O$.
This paper corrects a previous analysis of the $\langle TTO \rangle_{odd}$ correlator in \cite{Coriano:2023cvf}, where some non-zero solutions were missing.\\
Moving forward, one might wonder if there are other examples of scalar operators with $\Delta_3=0$ or 4 which generate non-vanishing $\langle JJO\rangle_{odd}$ and $\langle TTO\rangle_{odd}$ correlators. Moreover, it would also be interesting to determine if there is any sort of explanation for the non-vanishing solutions of the $\langle TTO\rangle_{odd}$ with $\Delta_3 = -2$ and 6.
We hope to revisit this topic in the future.\\
\vspace{0.5cm}\\
\centerline{\bf Acknowledgements}
This work is partially funded by the European Union, Next Generation EU, PNRR project "National Centre for HPC, Big Data and Quantum Computing", project code CN00000013; by INFN, iniziativa specifica {\em QG-sky} and by the grant PRIN 2022BP52A MUR "The Holographic Universe for all Lambdas" Lecce-Naples. 

\appendix

\section{3K Integrals} \label{appendix:3kint}
The most general solution of the CWIs for our correlators can be written in terms of integrals involving a product of three Bessel functions, namely 3K integrals. In this appendix, we will illustrate such integrals and their properties. For a detailed review on the topic, see also \cite{Bzowski:2013sza,Bzowski:2015pba,Bzowski:2015yxv}.
\subsection{Definition and properties}
First, we recall the definition of the general 3K integral
\begin{equation}\label{eq:3kintdef}
	I_{\alpha\left\{\beta_1 \beta_2 \beta_3\right\}}\left(p_1, p_2, p_3\right)=\int d x x^\alpha \prod_{j=1}^3 p_j^{\beta_j} K_{\beta_j}\left(p_j x\right)
\end{equation}
where $K_\nu$ is a modified Bessel function of the second kind 
\begin{equation}
	K_\nu(x)=\frac{\pi}{2} \frac{I_{-\nu}(x)-I_\nu(x)}{\sin (\nu \pi)}, \qquad \nu \notin \mathbb{Z} \qquad\qquad I_\nu(x)=\left(\frac{x}{2}\right)^\nu \sum_{k=0}^{\infty} \frac{1}{\Gamma(k+1) \Gamma(\nu+1+k)}\left(\frac{x}{2}\right)^{2 k}
\end{equation}
with the property
\begin{equation}
	K_n(x)=\lim _{\epsilon \rightarrow 0} K_{n+\epsilon}(x), \quad n \in \mathbb{Z}
\end{equation}
The 3K integral depends on four parameters: the power $\alpha$ of the integration variable $x$, and the three Bessel function indices $\beta_j$ . The arguments of the 3K integral are magnitudes of momenta $p_j$ with $j = 1, 2, 3$. One can notice the integral is invariant under the exchange $(p_j , \beta_j )\leftrightarrow (p_i , \beta_i )$.
We will also use the reduced version of the 3K integral defined as
\begin{equation}
	J_{N\left\{k_j\right\}}=I_{\frac{d}{2}-1+N\left\{\Delta_j-\frac{d}{2}+k_j\right\}}
\end{equation}
where we introduced the condensed notation $\{k_j \} = \{k_1, k_2, k_3 \}$.
The 3K integral satisfies an equation analogous to the dilatation equation with scaling degree
\begin{equation}
	\text{deg}\left(J_{N\left\{k_j\right\}}\right)=\Delta_t+k_t-2 d-N
\end{equation}
where 
\begin{equation}
	k_t=k_1+k_2+k_3,\qquad\qquad \Delta_t=\Delta_1+\Delta_2+\Delta_3
\end{equation}
From this analysis, it is simple to relate the form factors to the 3K integrals. Indeed, the dilatation Ward identity of each from factor tells us that this needs to be written as a combination of integrals of the following type
\begin{equation}
	J_{N+k_t,\{k_1,k_2,k_3\}}
\end{equation}
where $N$ is the number of momenta that the form factor multiplies in the decomposition.
Let us now list some useful properties of 3K integrals
\begin{equation}\label{eq:3kprop}
	\begin{aligned}
		&\frac{\partial}{\partial p_n} J_{N\left\{k_j\right\}}  =-p_n J_{N+1\left\{k_j-\delta_{j n}\right\}} \\&
		J_{N\left\{k_j+\delta_{j n}\right\}}  =p_n^2 J_{N\left\{k_j-\delta_{j n}\right\}}+2\left(\Delta_n-\frac{d}{2}+k_n\right) J_{N-1\left\{k_j\right\}} \\&
		\frac{\partial^2}{\partial p_n^2} J_{N\left\{k_j\right\}}  =J_{N+2\left\{k_j\right\}}-2\left(\Delta_n-\frac{d}{2}+k_n-\frac{1}{2}\right) J_{N+1\left\{k_j-\delta_{j n}\right\}}, \\&
		K_n J_{N\left\{k_j\right\}}  \equiv\left(\frac{\partial^2}{\partial p_n^2}+\frac{\left(d+1-2 \Delta_n\right)}{p_n} \frac{\partial}{\partial p_n}\right) J_{N\left\{k_j\right\}}=J_{N+2\left\{k_j\right\}}-2 k_n J_{N+1\left\{k_j-\delta_{j n}\right\}},\\&
		K_{n m} J_{N\left\{k_j\right\}}\equiv (K_n-K_m)J_{N\left\{k_j\right\}} =-2 k_n J_{N+1\left\{k_j-\delta_{j n}\right\}}+2 k_m J_{N+1\left\{k_j-\delta_{j m}\right\}}.
	\end{aligned}
\end{equation}
\subsection{Appell functions}
In the case of scalar primary operators, for example, of scaling dimensions $\Delta_i$, and momenta $p_1,p_2,p_3$,  the solutions expressed by 3K integrals can be directly related to the four Appell functions \cite{Coriano:2013jba} characterized by four pairs of indices $(a_i,b_j)$ $(i,j=1,2)$. These are the indices that in the change of variables $(p_1^2, p_2^2, p_3^2)\to 
(p_1^2,x,y)$, $x=p_2^2/p_1^2,$ $y=p_3^2/p_1^2$ reduce the special conformal constraints to Appell hypergeometric equations deprived of $1/x$ or $1/y$ singularities \cite{Coriano:2018bbe,Coriano:2018bsy}. 
Setting 
\begin{equation}
\alpha(a,b)= a + b + \frac{d}{2} -\frac{1}{2}(\Delta_2 +\Delta_3 -\Delta_1) \qquad \beta (a,b)=a +  b + d -\frac{1}{2}(\Delta_1 +\Delta_2 +\Delta_3) \qquad 
\label{alphas}
\end{equation}
we can introduce the following function
\begin{align}
\label{F4def}
F_4(\alpha(a,b), \beta(a,b); \gamma(a), \gamma'(b); x, y) = \sum_{i = 0}^{\infty}\sum_{j = 0}^{\infty} \frac{(\alpha(a,b), {i+j}) \, 
	(\beta(a,b),{i+j})}{(\gamma(a),i) \, (\gamma'(b),j)} \frac{x^i}{i!} \frac{y^j}{j!} 
\end{align}
where $(\alpha,i)=\Gamma(\alpha + i)/ \Gamma(\alpha)$ is the Pochhammer symbol. We will refer to $\alpha\ldots \gamma'$ as to the first,$\ldots$, fourth parameters of $F_4$.\\ 
The four independent solutions of the Appell system of equations are then all of the form $x^a y^b F_4$, linearly combined in a Bose-symmetric form. Specifically, the solution for the parity-even correlator with three scalar operators takes the general form 
\begin{equation}
\Phi(p_1,p_2,p_3)=p_1^{\Delta-2 d} \sum_{a,b} c(a,b,\vec{\Delta})\,x^a y^b \,F_4(\alpha(a,b), \beta(a,b); \gamma(a), \gamma'(b); x, y) 
\label{compact}
\end{equation}
where the sum runs over the four values $a_i, b_i$ $i=0,1$ with arbitrary constants $c(a,b,\vec{\Delta})$, with $\vec{\Delta}=(\Delta_1,\Delta_2,\Delta_3)$. Specifically, 
\begin{align}
&\alpha_0\equiv \alpha(a_0,b_0)=\frac{d}{2}-\frac{\Delta_2 + \Delta_3 -\Delta_1}{2},\, && \beta_0\equiv \beta(b_0)=d-\frac{\Delta_1 + \Delta_2 +\Delta_3}{2}, 
\end{align}
\begin{align}
&\gamma_0 \equiv \gamma(a_0) =\frac{d}{2} +1 -\Delta_2,\, & &\gamma'_0\equiv \gamma(b_0) =\frac{d}{2} +1 -\Delta_3
\end{align}
for a generic $d$ dimension.  

\subsection{Zero momentum limit} \label{appzerolimit3k}
When solving the secondary CWIs, it may be useful to perform a zero momentum limit.
In this subsection, we review the behavior of the 3K integrals in the limit $p_3\rightarrow0$. In this limit, the momentum conservation gives
\begin{equation}
	p_{1}^{\mu}=-p_{2}^{\mu} \qquad \Longrightarrow \qquad p_1=p_2 \equiv p
\end{equation}
Assuming that $\alpha>\beta_t-1$ and $\beta_3>0$, we can write
\begin{equation}
	\lim _{p_3 \rightarrow 0} I_{\alpha\left\{\beta_j\right\}}\left(p, p, p_3\right)=p^{\beta_t-\alpha-1} \ell_{\alpha\left\{\beta_j\right\}}
\end{equation}
where 
\begin{equation}
	\ell_{\alpha\left\{\beta_j\right\}}=\frac{2^{\alpha-3} \Gamma\left(\beta_3\right)}{\Gamma\left(\alpha-\beta_3+1\right)} \Gamma\left(\frac{\alpha+\beta_t+1}{2}-\beta_3\right) \Gamma\left(\frac{\alpha-\beta_t+1}{2}+\beta_1\right) \Gamma\left(\frac{\alpha-\beta_t+1}{2}+\beta_2\right) \Gamma\left(\frac{\alpha-\beta_t+1}{2}\right)
\end{equation}
\normalsize
We can derive similar formulas for the case $p_1\rightarrow0$ or $p_2\rightarrow0$ by 
considering the fact that 3K integrals are invariant under the exchange $(p_j , \beta_j )\leftrightarrow (p_i , \beta_i )$.

\subsection{Divergences and regularization}
The 3K integral defined in \eqref{eq:3kintdef} converges when 
\begin{equation}
	\alpha>\sum_{i=1}^3\left|\beta_i\right|-1 \quad ; \quad p_1, p_2, p_3>0
\end{equation}
If $\alpha$ does not satisfy this inequality, the integrals must be defined by an analytic continuation. 
The quantity
\begin{equation}
	\delta \equiv \sum_{j=1}^3\left|\beta_j\right|-1-\alpha
\end{equation}
is the expected degree of divergence.
However, when
\begin{equation}
	\alpha+1 \pm \beta_1 \pm \beta_2 \pm \beta_3=-2 k \quad, \quad k=0,1,2, \dots
\end{equation}
for some non-negative integer $k$ and any choice of the $\pm$ sign, the analytic continuation of the 3K integral generally has poles in the regularization parameter. 
Therefore, if the above condition is satisfied, we need to regularize the integrals. This can be done by shifting the parameters of the 3K integrals as
\begin{equation}
	I_{\alpha\left\{\beta_1, \beta_2, \beta_3\right\}} \rightarrow I_{\tilde{\alpha}\left\{\tilde{\beta}_1, \tilde{\beta}_2, \tilde{\beta}_3\right\}} \quad \Longrightarrow \quad J_{N\left\{k_1, k_2, k_3\right\}} \rightarrow J_{N+u \epsilon\left\{k_1+v_1 \epsilon, k_2+v_2 \epsilon, k_3+v_3 \epsilon\right\}}
\end{equation}
where
\begin{equation}
	\tilde{\alpha}=\alpha+u \epsilon \quad, \quad \tilde{\beta}_1=\beta_1+v_1 \epsilon \quad, \quad \tilde{\beta}_2=\beta_2+v_2 \epsilon \quad, \quad \tilde{\beta}_3=\beta_3+v_3 \epsilon
\end{equation}
or equivalently by considering
\begin{equation}
	d \rightarrow d+2 u \epsilon \quad ; \quad \Delta \rightarrow \Delta_i+\left(u+v_i\right) \epsilon
\end{equation}
In general, the regularization parameters $u$ and $v_i$ are arbitrary. However, in certain cases, there may be some constraints on them. 

\subsection{3K integrals and Feynman integrals}
3K integrals are related to Feynman integrals in momentum space. The exact relations were first derived in \cite{Bzowski:2013sza,Bzowski:2015yxv}. Here we briefly show the results. Such expressions have been recently used in order to show the connection between the conformal analysis and the perturbative one for the $\langle AVV\rangle$ correlator \cite{Coriano:2023hts}.\\
Let $K_{d\{\delta_1\delta_2\delta_3\}}$ denote a massless scalar 1-loop 3-point momentum space integral
\begin{equation}
	K_{d\left\{\delta_1 \delta_2 \delta_3\right\}}=\int \frac{\mathrm{d}^d \boldsymbol{k}}{(2 \pi)^d} \frac{1}{k^{2 \delta_3}\left|\boldsymbol{p}_1-\boldsymbol{k}\right|^{2 \delta_2}\left|\boldsymbol{p}_2+\boldsymbol{k}\right|^{2 \delta_1}}
\end{equation}
Any such integral can be expressed in terms of 3K integrals and vice versa. For scalar integrals the relation reads
\begin{equation}
	K_{d\{\delta_1\delta_2\delta_3\}}=\frac{2^{4-\frac{3d}{2}}}{\pi^{\frac{d}{2}}}\times\frac{I_{\frac{d}{2}-1\{\frac{d}{2}+\delta_1-\delta_t,\frac{d}{2}+\delta_2-\delta_t,\frac{d}{2}+\delta_3-\delta_t \}}}{\Gamma(d-\delta_t)\Gamma(\delta_1)\Gamma(\delta_2)\Gamma(\delta_3)}
\end{equation}
where $\delta_t=\delta_1+\delta_2+\delta_3$. Its inverse reads
\begin{equation}
	\begin{aligned}
		& I_{\alpha\left\{\beta_1 \beta_2 \beta_3\right\}}=2^{3 \alpha-1} \pi^{\alpha+1} \Gamma\left(\frac{\alpha+1+\beta_t}{2}\right) \prod_{j=1}^3 \Gamma\left(\frac{\alpha+1+2 \beta_j-\beta_t}{2}\right) \\
		&\hspace{2.5cm} \times K_{2+2 \alpha,\left\{\frac{1}{2}\left(\alpha+1+2 \beta_1-\beta_t\right), \frac{1}{2}\left(\alpha+1+2 \beta_2-\beta_t\right), \frac{1}{2}\left(\alpha+1+2 \beta_3-\beta_t\right)\right\}}
	\end{aligned}
\end{equation}
where $\beta_t=\beta_1+\beta_2+\beta_3$. All tensorial massless 1-loop 3-point integrals can also be expressed in terms of a number of 3K integrals when their tensorial structure is resolved by standard methods (for the exact expressions in this case see Appendix A.3 of \cite{Bzowski:2013sza}).

\providecommand{\href}[2]{#2}\begingroup\raggedright\endgroup

\end{document}